\documentstyle[12pt]{report}
\begin{document}
\begin{center}
{\large{\bf Systematic Study of Shell Effect near Drip-lines}}
\vskip 1cm
S. Adhikari$^a$ and C. Samanta$^{a,b}$
\vskip 0.2cm {\it $^a$Saha
Institute of Nuclear Physics,\\
 1/AF Bidhannagar, Kolkata 700 064}\\
and\\
{\it $^b$Physics Department, Virginia Commonwealth University\\
Richmond, Virginia 23284-2000}\\
\end{center}
\vskip 0.2cm

\noindent Variation of nuclear shell effects with nucleon numbers 
are evaluated using the modified Bethe-Weizs$\ddot{a}$cker 
mass formula (BWM) and the measured atomic masses.
The shell effects at magic neutron numbers N=8, 20, 28, 50, 82 and 126  and  
magic proton numbers Z=8, 20, 28,50 and 82 are found to vary rapidly approaching
the drip lines. The shell effect increases when approaches another 
magic number. Thus, shell effects are not always negligible near the
drip lines.\\ 

\vskip 0.5cm


\noindent With increasing discoveries of more and more nuclei away from the 
valley of stability, the domain of nuclear-magicity has started changing. 
In neutron (N) or, proton (Z)-rich nuclei unusual stabilities are found
at various N or, Z values which are very different from the well known magic
numbers 2, 8, 20, 28, 50, 82 and 126 $\cite{{oz00},{sa02}}$. On 
the other hand, loss of magicity at N=8 (Z=4) $\cite{sa02}$ has been 
experimentally confirmed $\cite{na00}$. Quenching or, loss of magicity in 
nuclei away from the valley of stability 
is a topic of great current interest because of its direct relevance to the 
production of heavy elements both in the laboratory and at the astrophysical 
sites. Chen et al. showed $\cite{ch95}$ that theoretical 
calculations incorporating quenching of 
the N=82 shell gap for Z$<50$ nuclei leads to a filling of the 
known abundance troughs arround A 
$\sim$ 120 and 140 and generates a better overall reproduction  of the heavy 
elements.  However, there are still some discrepancies 
which calls for further investigation.
As quenching of magic shell gap near n- and p-drip line would have 
significant effect on the astrophysical processes, a systematic 
study of the possible quenching of the magic shell effects 
away from the valley of stability is essential. In this work we present 
a systematic study of the possible change 
of shell effects for nucleon number 8, 20, 28, 50, 82 and N=126 through a 
comparison of experimental data and results from a mass formula without shell 
effect.\\

\noindent  A mass formula based on the liquid drop model 
was first prescribed by Bethe-Weizs$\ddot{a}$cker (BW) \cite{{kr--},{hy--},{pr--}}. It  
was designed to fit the 
heavy and medium mass nuclei. As it has no shell correction incorporated
the BW fails near the magic numbers where the closed shell structure of nuclei 
demands extra stability. For years this inadequacy has been utilized 
to identify the magic numbers as they stand out as marked deviations 
\cite{{kr--},{hy--},{pr--}}. With the 
discovery of nuclei away from the valley of stability it was noticed that the 
BW formula is inadequate near the drip lines, especially for light nuclei $\cite{sa02}$. 
The improved
liquid drop model (ILDM) $\cite{so03}$ also does not reproduce the trend of the
binding energy versus neutron number curves of light nuclei correctly. Earlier we suggested
a modified 
Bethe-Weizs$\ddot{a}$cker mass formula (BWM) in which the binding energy is defined
in terms of mass number A (=N+Z) and proton number (Z) 
as $\cite{sa02}$,

\begin{eqnarray}
BE(A,Z)=15.777 A-18.34 A^{2/3}-0.71\frac{Z(Z-1)}{A^{1/3}}-\frac{23.21}
{(1+e^{-A/17})}\times\frac{(A-2Z)^2}{A}
\nonumber \\
+ (1 - e^{-A/30}) \delta,
\end{eqnarray}

\noindent where,  $\delta$ = +12 A$^{-1/2}$
for even Z-even N nuclei, and -12 A$^{-1/2}$ for odd Z-odd N nuclei and 0
for odd A nuclei. The BWM reproduces the general trend of the binding energy 
versus neutron number curves for all nuclei from Li
to Bi. Like BW, the BWM also does not contain any shell correction or 
Wigner term $\cite{wi37}$. Therefore, it overpredicts the mass near the magic
numbers as well as for nuclei with N=Z. As the shell effect quenches, the discrepancy between the experimental 
data and the 
predictions of BWM diminishes. Therefore BWM can be used to identify 
extrastability 
as well as, quenching of shell effect through a comparison with the 
experimental mass data.
 In the following, we study the change of shell effects with nucleon numbers
 near the magic numbers
8, 20, 28, 50, 82 and 126  using the BWM
and experimental mass data \cite{{wa00},{no02},{sc01}}. \\

\noindent The one nucleon separation energy of a nucleus $(A,Z)$ is defined as, 
\begin{eqnarray}
S_n(A,Z)&=&BE(A,Z) - BE(A-1,Z)\\
S_p(A,Z)&=&BE(A,Z) - BE(A-1,Z-1)
\end{eqnarray}

\noindent When this S$_n$ or, S$_p$ are plotted against N or Z, a large drop 
is found after the N or Z values which correspond to magic numbers. BWM 
cannot reproduce this break as it has no shell effect incorporated. 
Fig.1(a) and Fig. 1(b) show one-neutron separation energy (S$_n$) versus neutron 
number plots for Be(Z=4) and O(Z=8). For O(Z=8) (Fig.1b), 
a large break at N=8 is clearly seen in experimental data which is not 
reproduced by BWM. 
On the otherhand, for Z=4 there is no large break in the experimental data 
at N=8 and, the difference between the BWM and experimental data almost
disappears. This indiactes loss of N=8 magicity in Z=4 nucleus and it has been 
experimentally confirmed $\cite{na00}$. Incidentally a large break can be seen 
in the experimental data at N=4 for Be(Z=4), which arises due to extrastability
 of N=Z nuclei. BWM cannot reproduce this large break as Wigner term is not 
incorporated in BWM. Similar zones of quenching of 
shell effects can be seen by plotting the one-nucleon separation energy 
derived from experimental data and the BWM predictions \cite{DNB04}. The 
difference ($\Delta$B) between the experimental binding energies (BE(EXP))
and theoretical (BE(BWM)) ones are computed using 
experimental masses \cite{{wa00},{no02},{sc01}} and the mass formula
BWM. 
In Fig.1(d) it can be seen that $\Delta$B has a large value at neutron magic number N=8 
whereas, no such peak can be seen at N=8 for Be (Fig. 1(c)). This delineates
the expected magicity at N=8 for Z=8 and,
the loss of N=8 magicity at Z=4. In Fig. 1(c) the extrastability
at N=Z=4 and at magic number N=2 are clearly seen. The new magic number N=16
\cite{oz00,sa02} also emerges clearly (Fig. 1(d)) in this comparison with 
the experimantal data and BWM.\\

\noindent The two nucleon separation energy (S$_{2i}$, i=n,p), the socalled, 
"shell gap" (G$_{2i}$) and the "shell effect" ($\Delta$B) of a nucleus (A,Z) are 
defined as, 
\begin{eqnarray}
S_{2n}(A,Z)&=&BE(A,Z) - BE(A-2,Z)\\
S_{2p}(A,Z)&=&BE(A,Z) - BE(A-2,Z-2)\\
G_{2n}(A,Z)&=&S_{2n}(A,Z) - S_{2n}(A+2,Z)
\nonumber \\
&=&2BE(A,Z) - BE(A-2,Z) - BE(A+2,Z)\\
G_{2p}(A,Z)&=&S_{2p}(A,Z) - S_{2p}(A+2,Z+2)
\nonumber \\
&=&2BE(A,Z) - BE(A-2,Z-2) - BE(A+2,Z+2)\\
\Delta B&=&BE(EXP) - BE(BWM)
\end{eqnarray}
\noindent The G$_{2i}$ is usually 
plotted with the experimental mass data \cite{{wa00},{no02},{sc01}} to monitor 
the change of shell gap, but it can not be used if there is a drastic 
change of shape in nuclei associated with the evaluation of G$_{2i}$. Whereas,
the $\Delta$B being simply the difference between the experimental and theoretical
mass is not affected by the deformation of other nuclei. \\

\noindent Plots of G$_{2n}$ 
versus proton number for N=8, 20, 28, 50, 82 and 126 are shown in Fig.2 . 
Similar plots of G$_{2p}$ versus neutron number 
for Z=8, 20, 28, 50 are presented in 
Fig.3. The mass formula BWM, which has no shell correction
or, Wigner effect incorporated, predicts 
a smooth continuous line and thus acts as a base line for comparison. 
There are several interesting features in 
the G$_{2i}$ plots. The so called "shell gap"  of all these 
magic numbers appears to 
change with nucleon numbers but they peak at N or, Z values where they get 
support from another magic number. The extrastability is found to decrease 
on both sides of the main peak. In Fig.2 a rise in experimental G$_{2n}$ 
values above the BWM predictions is seen for N=8(Z=7-9), N=20(Z=10, 15-17, 19-21),
N=28(Z=15, 16, 18-24, 27-32). The shell effects for N=50, 82 and 126 do not disappear
in the existing experimental data. For N=28, three peaks are seen at Z=16, 20, 28 of 
which the last two are known magic numbers. 
In Fig.3 similar rise is seen for Z=8(N=7-9, 13-16), Z=20(N=19-21, 27-31, 33).
For Z=28, BWM overpredicts only at N=25 and for Z=50, the experimental 
G$_{2p}$ values are always far above BWM.
A comparison with the predictions of BWM indicates 
strong quenching of G$_{2n}$ at N=8(Z=4-6, 10), 
N=20(Z=11-14, 18, 22-24), N=28(Z=17, 25, 26, 30). Some quenching is seen at 
N=50(Z=32), N=82(Z=49), N=126(Z=81, 80, 83-90). Similar quenching of 
G$_{2p}$ can be seen for Z=8(N=6, 10-12), Z=20(N=16-18, 22-26), Z=28(N=25).\\ 

\noindent The plot of G$_{2p}$ for Z=82 is presented in Fig.4 along with 
predictions of different mass formulae \cite{{sa02},{sn83},{ko00},{mo95},{so03}}. 
The exerimental G$_{2p}$ value queches on both sides of the peak at N=126. 
At $N=106$ the difference between experimental value and BWM 
prediction almost vanishes ($\sim$ 140 keV) but, increases again at 
N$<$106. Interestingly, the recent 
mass formula of Koura et al. \cite{ko00}, which is known to be valid for both
light and heavy nuclei, reproduces most of the details of 
the experimental data for G$_{2i}$ distributions of lower magic numbers (Fig.2 and 3). 
But for heavier nuclei it delineates significant overprediction near 
drip lines and under prediction near the peaks.
Results from the mass formula of M$\ddot{o}$ller et al. \cite{mo95}, 
Satpathy-Nayak \cite{sn83} and ILDM \cite{so03} are also enclosed 
for comparison. Although they have shell correction incorporated, none of them
reproduces the exact nature of the G$_{2p}$ distribution of Z=82.\\

\noindent In Fig.5, the difference ($\Delta$B) between the binding
energies computed from the experimental masses $\cite{{wa00},{no02},{sc01}}$
and the BWM  is plotted against the proton numbers for neutron magic
nuclei. Similar plot for the proton magic nuclei are presented in Fig. 6
against the neutron numbers.
The BWM being basically a liquid drop model without shell effect acts as the base line for
evaluation of the shell effect in nuclei. From these figures it is clear that the 
shell effect does not always quench near the drip lines. 
In some nuclei the shell effect,
after a quenching near the mid shell region, actually increases near the drip line
as another magic number is approached. In Fig. 5, for  N=8, the extrastabilty
disappears for Z=3-5. It is interesting to note that in Fig. 5, N=28 shows
extrastability at Z=28, 20 and a mild increase towards Z=16
and, the fall after Z=28 is rather flat in the region Z=31-33. 
In Fig. 6, for Z=16, the extrastability does not reduce
after the magic neutron number N=28. On the contrary, at N=32 it is even higher. 
For Z=20 in Fig. 6, the N=29, 31 and 33 have higher $\Delta$B values than N=28.
This suggests that the neutron magic number in this region might be at higher N values.
New magicities at N=30 and 32 have been predicted earlier around Z=20 region \cite{ri02}.
However, as some of these data points are from systematics only,
additional mass measurement for neutron-rich Sulfur isotopes 
 are needed to confirm this probable shift of neutron magicity.\\

\noindent From Figs. 3, 4 and
Figs. 5, 6 it is clear that quenching or, loss of shell effects or, extrastability 
 demonstrated by $\Delta$B and
G$_{2p}$ are not always the same. The reason for this anomaly is the way they are evaluated.
The evaluation of
G$_{2p}$ at each point involves masses of three nuclei. Therefore, its use as a measure of 
shell gap is restricted if there is a  drastic change in shape
or, deformation in any of the three nuclei. The $\Delta$B gives the  
measure of the shell effect (not the shell gap), and it involves only 
the mass of one nucleus at each point.\\

\noindent Recently, experimental evidence of N=82 shell
quenching has been claimed by observing the high Q$_\beta$ value
for $^{130}$Cd $\cite{di03}$ which can not be explained by the "unquenched" 
finite-range droplet model(FRDM). In one-neutron separation energy versus
N plot (Fig.7) one can see the large break
at N=82 for the magic nucleus Sn (Z= 50) which can not be reproduced by BWM, as expected.
Similar break in Te (Z=52) is seen at N=82, 
but with a lesser magnitude. Like the Sn and Te, the S$_n$ versus N data for Cd also
shows a gradually increasing discrepency between the experimental data and 
BWM as it approaches the magic number N=82 beyond which no experimental
mass data is available so far. It only shows a large break at N=50
indicating existence of large shell effect at N=50 for Z=48. 
Measurements of the mass of $^{129,131,132}$Cd are essential to see whether
there is a break in S$_n$ after N=82 and how big it is. In view of the predicted quenching of 
N=82 shell gap in Cd, a considerable reduction in this break is expected.\\ 

\noindent In summary the change of shell effects in nuclei with N or Z=8, 20, 28, 
50, 82 and N=126 approaching the drip lines have been studied. A modified 
Bethe-Weisz$\ddot{a}$cker mass formula (BWM)$\cite{sa02}$ is employed for comparison. 
The BWM reproduces the general trend of the binding energy versus nucleon number 
curves for Li to Bi nuclei much better than the existing macroscopic formulas. It
has no shell effect incorporated. If the shell effect quenches, the discrepancy 
between the BWM predictions and experimental data diminishes. 
Thus BWM serves as a baseline for comparison.
The G$_{2i}$ and $\Delta$B vales are used here to monitor the change of 
shell effects with change of neutron and proton numbers.
Mass formulae of Koura et al. $\cite{ko00}$,  M$\ddot{o}$ller et al. 
$\cite{mo95}$ and Satpathy-Nayak $\cite{sn83}$, which have built in shell effects,
are used for comparison. The shell effects are found to vary with the
nucleon numbers and increases if it approaches another magic number. 
Several domains of  
shell effect quenching and extrastability 
are found for Z=4, 6, 8, 16, 20, 28, 82 and N=6, 8, 16, 20, 28, 82 and 126 of 
which loss of magicity at N=8(Z=4) shown here is already confirmed experimentally.
Further experimental data are needed to confirm many of the observations
suggested by this work.

\newpage
\noindent{\bf Figure Caption :} 

\noindent {\bf Fig. 1(a),(b)} Plots of one-neutron separation energy(S$_n$) 
versus neutron number(N) 
curve from experimental data $\cite{wa00}$ and BWM to delineate strength of 
shell effect at N=8 for Z=4, 8; {\bf (c),(d)} plots of $\Delta$B (=BE(EXP)-BE(BWM))
versus neutron 
number(N) curve from experimental data $\cite{wa00}$ and BWM to delineate 
strength of shell effect at N=8 for for Z=4, 8.

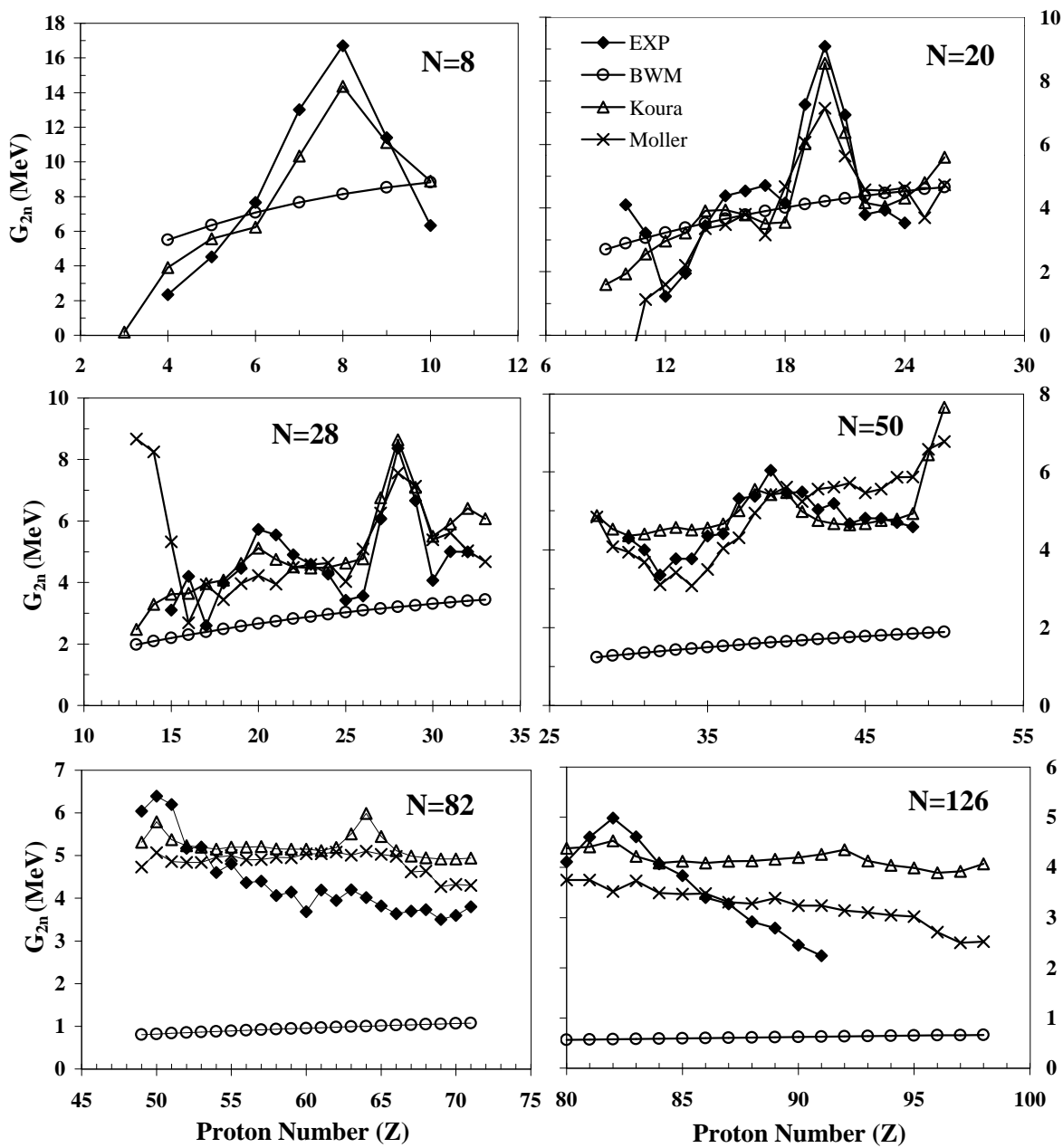
\noindent {\bf Fig. 2} Plots of G$_{2n}$ = S$_{2n}$(N) - S$_{2n}$(N+2)
versus proton number (Z), computed
from measured masses $\cite{{wa00},{no02}}$ and from the 
mass formulas of BWM $\cite{sa02}$, Koura et al. 
$\cite{ko00}$, and M$\ddot{o}$ller et al. $\cite{mo95}$  
for magic neutron numbers N=8, 20, 28, 50, 82 and 126. 

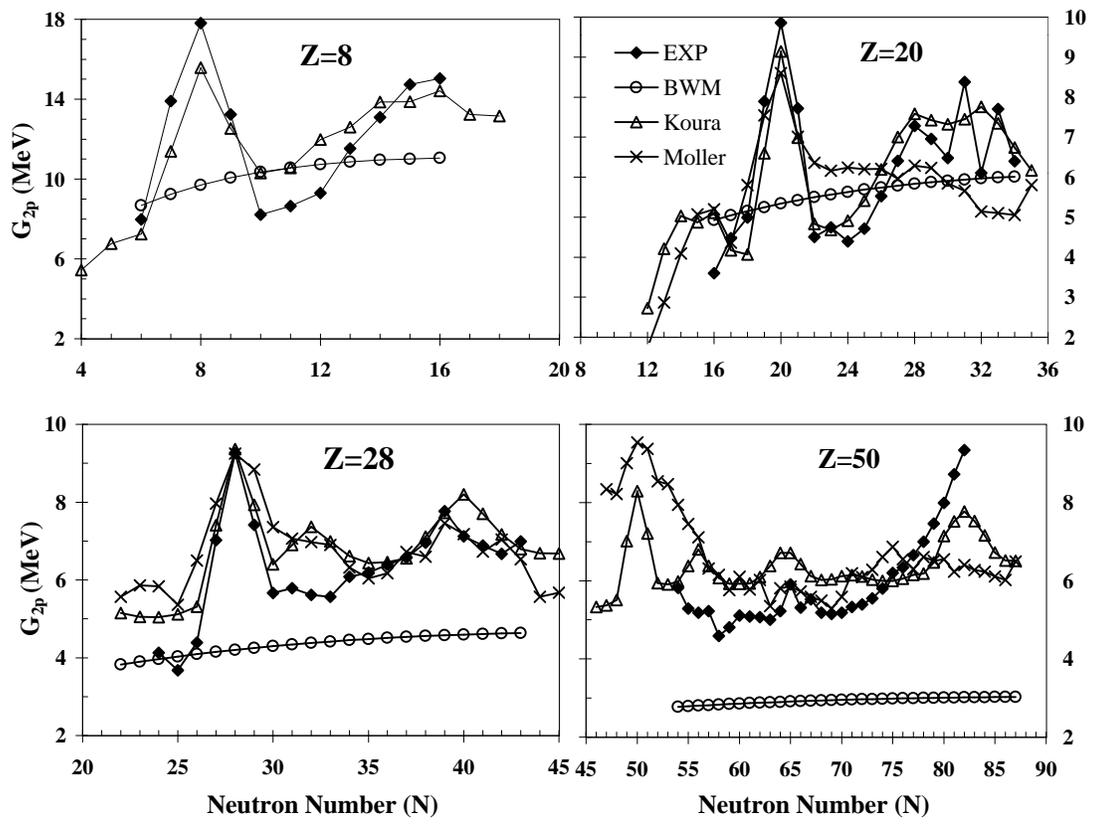
\noindent {\bf Fig. 3} Plots of G$_{2p}$ = S$_{2p}$(Z) - S$_{2p}$(Z+2) versus
neutron number, computed 
from measured masses $\cite{{wa00},{no02}}$ and  from the mass 
formulas of BWM  $\cite{sa02}$, Koura et al. 
$\cite{ko00}$, and M$\ddot{o}$ller et al. $\cite{mo95}$ 
for magic proton numbers Z=8, 20, 28, 50.

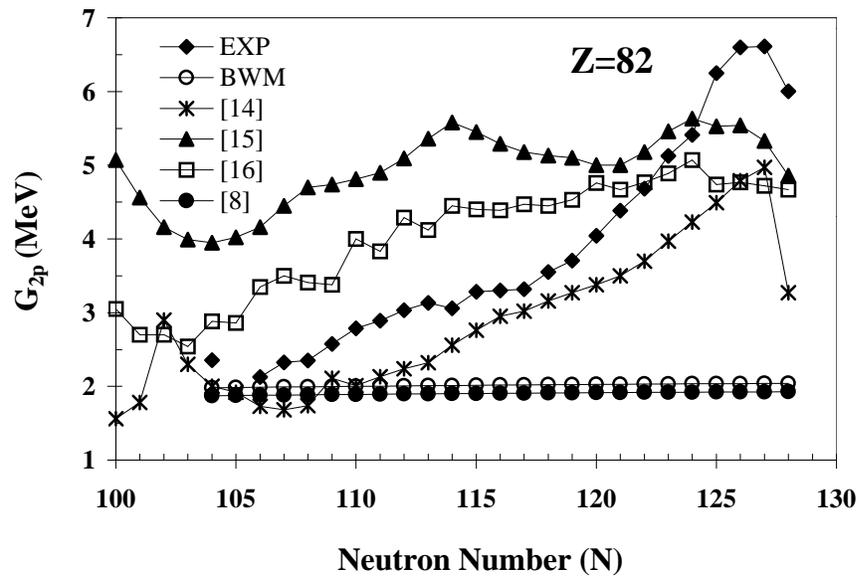
\noindent {\bf Fig. 4} Plots of G$_{2p}$ 
from measured mass $\cite{{wa00},{no02}, {sc01}}$ 
and from the mass formula of BWM $\cite{sa02}$,
Satpathy-Nayak \cite{sn83}, Koura et al.  $\cite{ko00}$, 
and M$\ddot{o}$ller et al. $\cite{mo95}$ 
for magic proton number  Z=82.

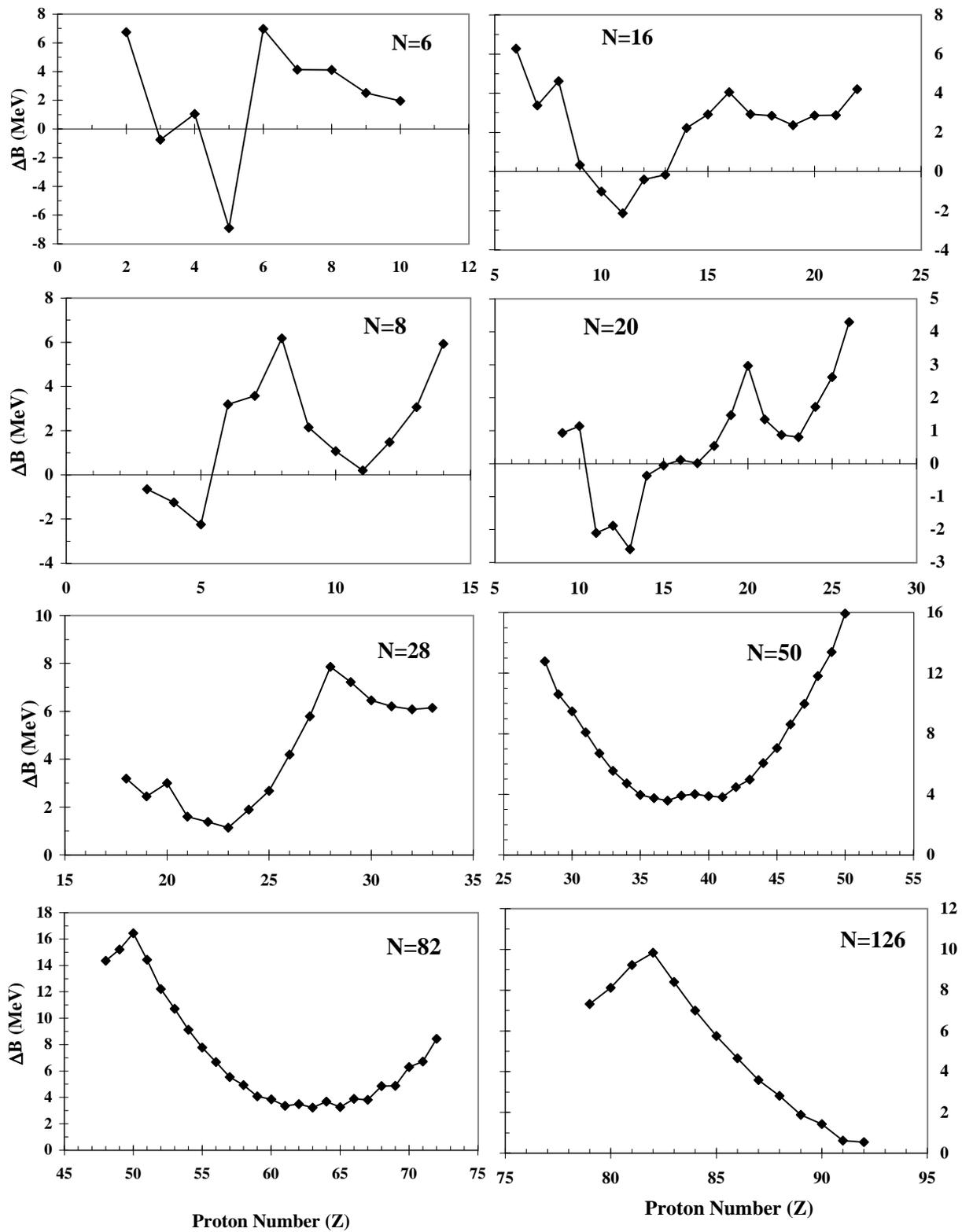
\noindent {\bf Fig. 5} Plots of $\Delta$B versus proton
number(N) from experimental data $\cite{{wa00},{no02},{sc01}}$ and BWM 
to show the variation of shell effects for N=6, 8, 16, 20, 28, 50, 82 and 126.

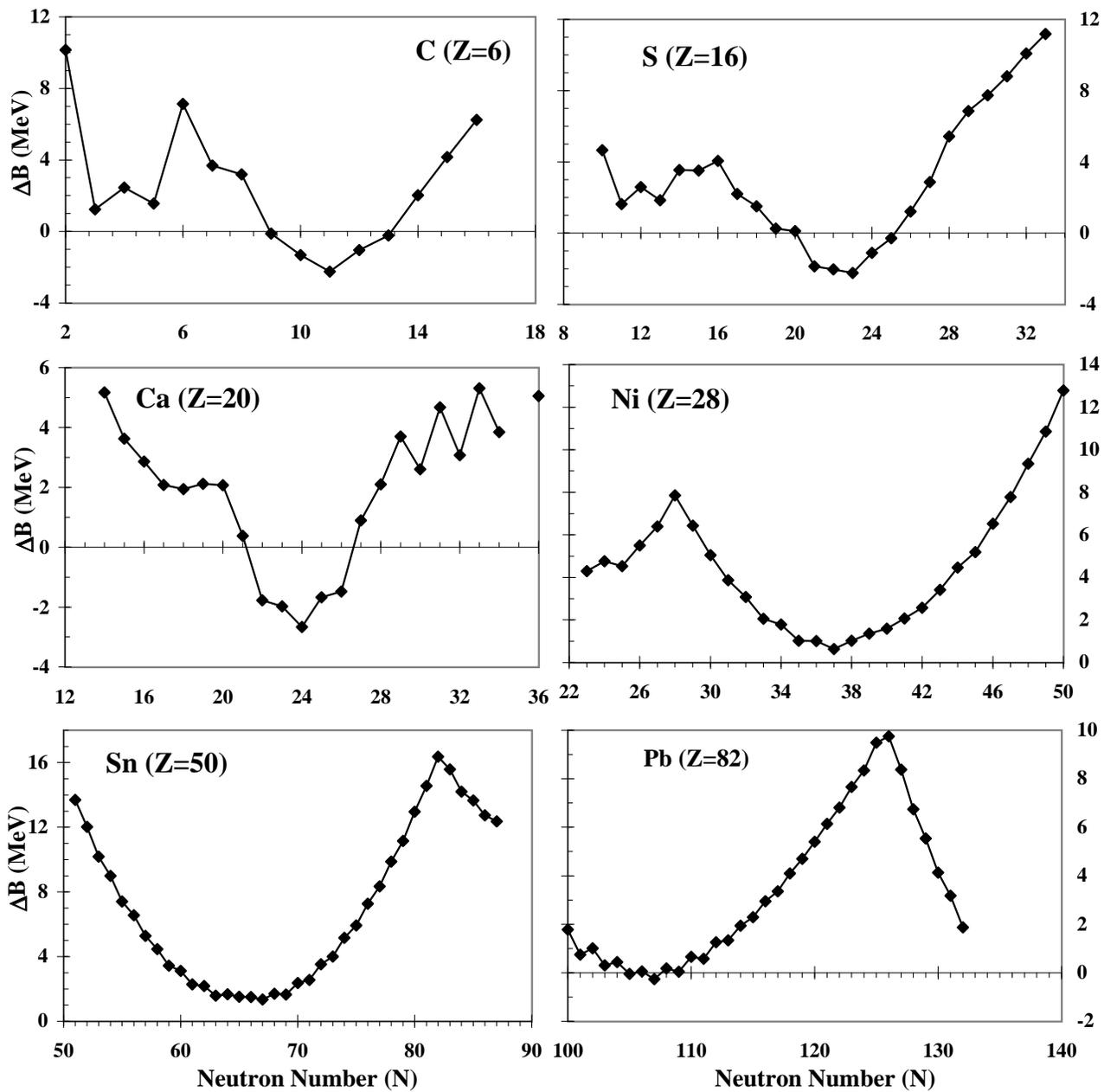
\noindent {\bf Fig. 6} Plots of $\Delta$B versus neutron
number(N) from experimental data \cite{{wa00},{no02},{sc01}} and BWM 
to show the variation of shell effects for Z=6, 16, 20, 28, 50, 82.

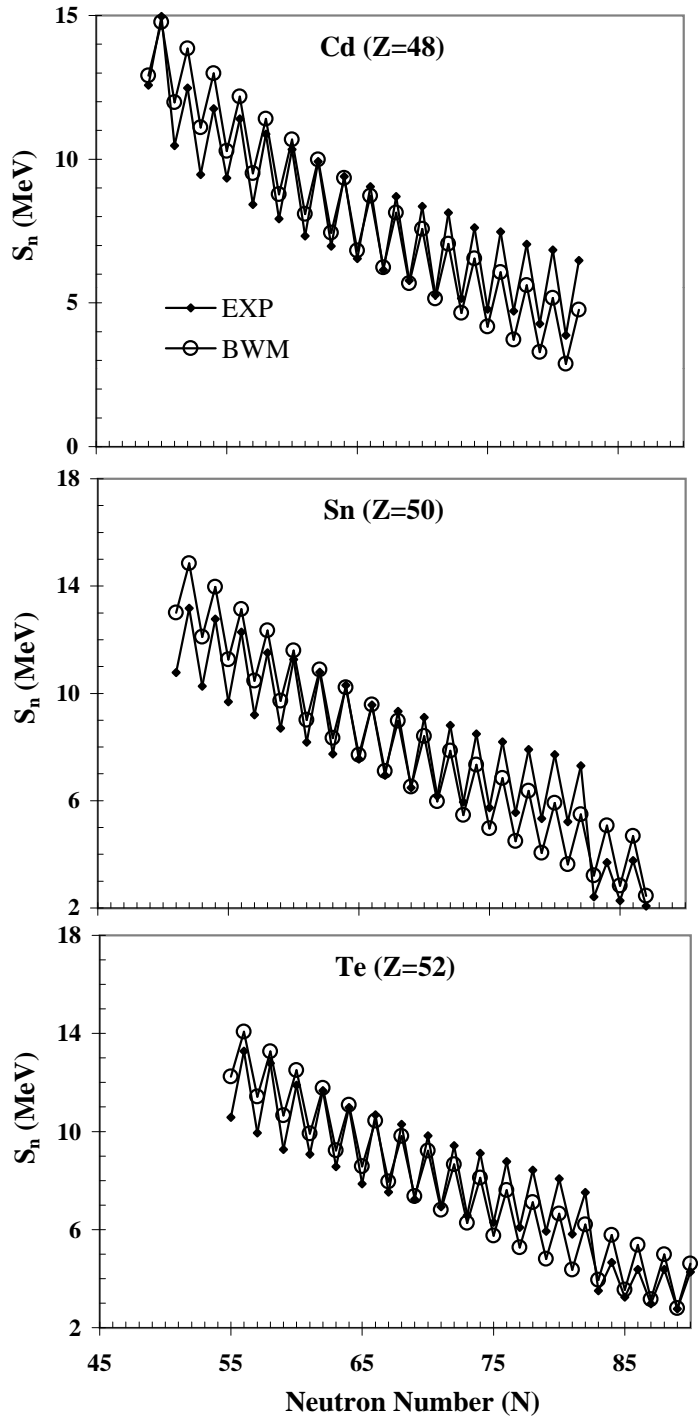
\noindent {\bf Fig. 7} Plots of one-neutron separation energy(S$_n$)
versus neutron number(N)
curve from experimental data $\cite{wa00}$ and BWM for Cd(Z = 48),
S$_n$(Z= 50) and, Te(Z= 52).  
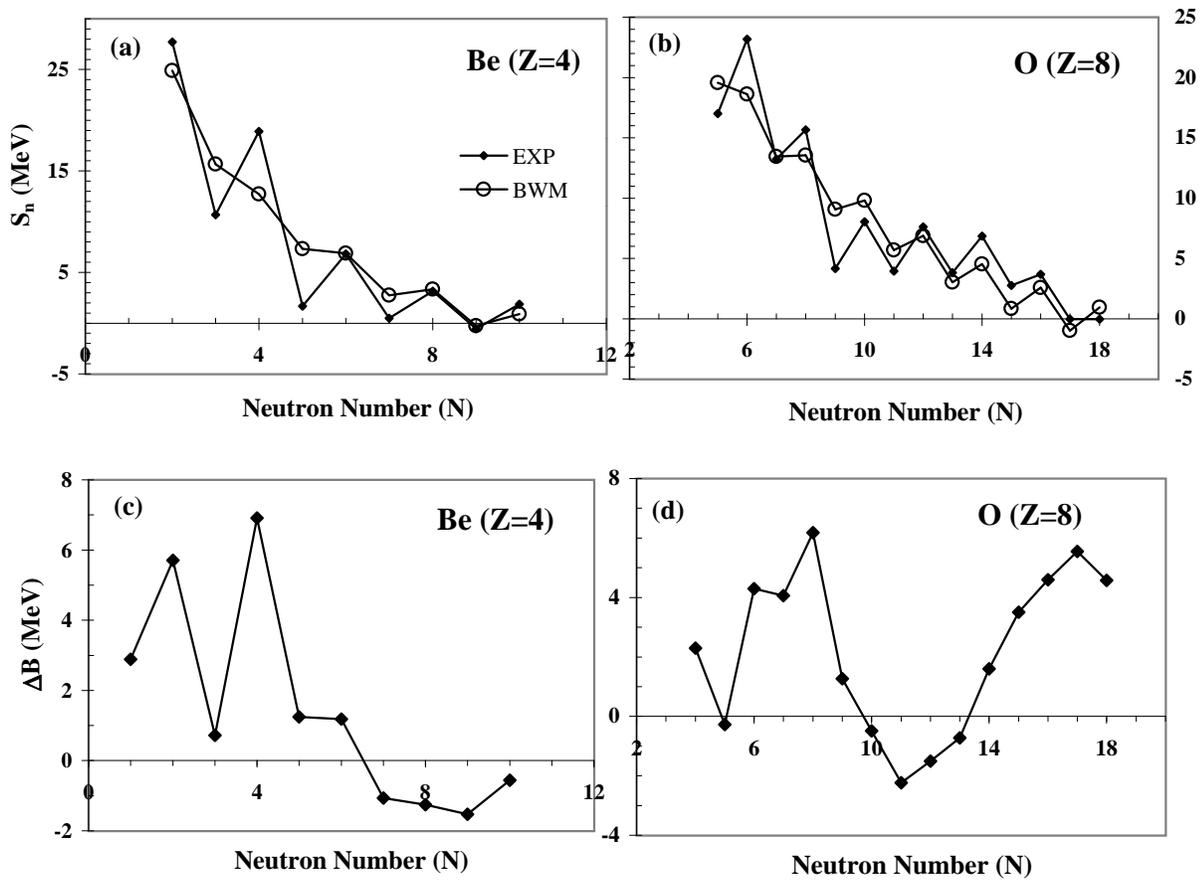
\end{document}